\def\ra{\rightarrow}
   \def\unlock{\catcode`@=11}

   \unlock

   \def\gsim{\mathrel{\mathpalette\@versim>}}

   \def\@versim#1#2{\vcenter{\offinterlineskip
        \ialign{$\m@th#1\hfil##\hfil$\crcr#2\crcr\sim\crcr } }}

\def\ie{{\elevenit i.e.}}
\def\hc{{H^+}}
\def\tt{{t\bar t}}
\def\bold#1{\setbox0=\hbox{$#1$}%
     \kern-.025em\copy0\kern-\wd0
     \kern.05em\copy0\kern-\wd0
     \kern-.025em\raise.0433em\box0 }
\documentstyle[12pt]{article}

\font\elevenbf=cmbx10 scaled\magstep 1
\font\elevenrm=cmr10 scaled\magstep 1
\font\elevenit=cmti10 scaled\magstep 1

\font\ninerm=cmr9

\renewenvironment{thebibliography}[1]
 { \elevenrm
   \begin{list}{\arabic{enumi}.}
    {\usecounter{enumi} \setlength{\parsep}{0pt}
     \setlength{\itemsep}{3pt} \settowidth{\labelwidth}{#1.}
     \sloppy
    }}{\end{list}}
\renewcommand{\thefootnote}{\fnsymbol{footnote}}

\begin{document}
\begin{titlepage}
%\vbox to 2cm{}
\hfill SLAC-PUB-6354

\hfill SCIPP 93/29

\hfill September 1993

\hfill T/E
\vglue 2.5cm
\begin{center}{{\elevenbf ENERGY DISTRIBUTION OF THE LEPTONS\\
                IN TOP DECAY TO CHARGED HIGGS\footnote{\ninerm
\baselineskip=9pt Work supported by the Department of Energy,
contract DE-AC03-76SF00515.}                \\}
\vglue 1.0cm
{\elevenrm ALEX POMAROL \\}

{\elevenit Santa Cruz Institute for Particle Physics\\}
{\elevenit University of California, Santa Cruz, CA 95064, USA\\}
\vglue 0.3cm
{\elevenrm and\\}
\vglue 0.3cm
{\elevenrm CARL R. SCHMIDT \\}
{\elevenit Stanford Linear Accelerator Center\\}
{\elevenit Stanford University, Stanford, CA 94309, USA\\}
\vglue 1.5cm
{\elevenrm ABSTRACT}}
\end{center}
\vglue 0.3cm
{\rightskip=3pc
 \leftskip=3pc
 \noindent
We analyze the
energy distribution of the leptons coming from the top decay
$t\ra  b\hc,\ bW^+
 \ra bl^+\nu$. The correlation between the lepton energy distribution
and the top spin
   may be useful for distinguishing the
$\hc$ signature from the $W$ background.
Such a correlation  can also be useful for disentangling
 the two different
$\hc \bar tb$ couplings
and determining
the experimental value of $\tan\beta$.
\vglue 0.6cm}
\vfill

\centerline{\it To appear in the Proceedings of the Workshop on}
\centerline{\it Physics at Current Accelerators and the Supercollider}
\centerline{\it Argonne National Laboratory, Argonne, Illinois}
\centerline{\it June 2-5, 1993}
\vglue 0.5cm
\pagebreak
\end{titlepage}
\elevenrm
\setcounter{page}{2}
\setcounter{footnote}{0}
\renewcommand{\thefootnote}{\arabic{footnote}}

%\noindent
Charged Higgs bosons ($H^+$) are present in models
with  two scalar doublets or a
 more extended Higgs sector \cite{gun}.
One notorious example
is the minimal supersymmetric model (MSSM).
In a general two Higgs doublet model (THDM)
the charged Higgs mass is a free parameter. In the MSSM, however,
 it is  bounded from below,
  $m_\hc\gsim m_W$.
Experimentally, the LEP Higgs search \cite{lep}
provides  a model-independent lower
 bound,
$m_\hc\gsim m_Z/2$. The recent experimental \cite{cleo}
 upper bound on the decay
$b\rightarrow s\gamma$ also constrains  $m_\hc$ but in
a model-dependent way \cite{bsg}.
 For example, in a THDM where one Higgs
doublet is  responsible for generating the masses of the down-type
fermions and the other Higgs doublet generates those of
the up-type (model II of ref.~\cite{gun}) we have $m_\hc\gsim 300$
GeV. Although the MSSM belongs to this type of model,  this bound does
not apply to it because one also has to consider
 the contribution  of the superpartners \cite{bsg}.

If $m_\hc<m_t-m_b$, the top quark  can decay to $\hc b$
providing a source of charged Higgs bosons.
The $t\ra\hc b$ branching ratio
 depends on the $\hc\bar tb$ coupling which is given
by (model II)
\begin{equation}
\frac{ig}{2\sqrt{2}m_W}\hc\bar t\left[m_t\cot\beta(1-\gamma_5)+
m_b\tan\beta(1+\gamma_5)\right]b\ +\ {\it h.c.} \, ,\label{coup}
\end{equation}
 where $\tan\beta$ is the
ratio of the vacuum expectation values of the two Higgs doublets.
{}From eq.~(\ref{coup}), we have
\begin{equation}
\Gamma(t\ra\hc b)=\frac{G_Fm_t}{8\sqrt{2}\pi}\left(1
-\frac{m_{\hc}^2}{m^2_t}\right)^2
\left[m^2_t\cot^2\beta+m^2_b\tan^2\beta\right]\, ,
\label{wi2}
\end{equation}
where we have neglected all purely kinematic factors
of $m_b$ that are not enhanced by $\tan\beta$.
Hence, the width of $t\ra\hc b$ is large and comparable to the standard
decay mode, $t\ra W^+b$, for large or small values of $\tan\beta$, \ie,
$\tan\beta< 1$ or $\tan\beta> 20$. In fig.~1, we show
BR($t\ra\hc b$)/BR($t\ra W^+ b$) as a function of $\tan\beta$ for
different values of
the charged Higgs mass.
 We can see that the BR
has a minimum at $\tan\beta\sim 5$. For $\tan\beta$  around
this value
 (this region,
$1<\tan\beta<20$,  is
  favored by low-energy supersymmetric models), the
 detection of $\hc$ may be difficult \cite{god}--\cite{barn}.
 Even  if the $t\ra\hc b$ branching ratio  and $m_\hc$
can be measured, there will still be an ambiguity in the value of
$\tan\beta$, \ie, $\tan\beta$  is a double-valued function of
 BR($t\ra\hc b$).

%\begin{figure}
%\plotpicture{\hsize}{8cm}{br.topdraw}
%\vskip-4cm
%\epsfysize=16cm
%\centerline{\epsffile{br.ps}}
%\vskip-4cm
%\vskip6pt
%\baselineskip=12pt
%\centerline{{\tenbf Figure 1:}  \tenrm
%Ratio between the decay $t\rightarrow H^+b$ and $t\rightarrow W^+b$.}
%\end{figure}

In this paper we will compute the energy distribution of the leptons
coming from the decay
 $t\ra b\hc\ra bl^+\nu$ and compare to that of leptons from
the decay $t\ra bW^+\ra bl^+\nu$ for polarized top quarks.
Expected differences in these distributions can be useful for
 distinguishing between charged Higgs signals and  $W$ signals.
 Furthermore, the lepton energy distribution from $t\ra b\hc\ra bl^+\nu$
 is sensitive to the
two terms of the coupling (\ref{coup}) and  will help resolve the
ambiguity in  $\tan\beta$.

The idea is based on the correlation between the lepton energy and
the top helicity \cite{cza}:
The decay $t\ra W^+b$ is dominated by a V--A interactions, \ie,
the $b$ is always left-handed. If the top mass is large,
the $W$ is longitudinal and
a $t_L$  will decay to an energetic $b_L$ (since it has to go forward
to carry the top spin)
and to a less-energetic $W$ that will decay into a less-energetic
lepton. Similarly, a $t_R$ will tend to decay to a
more-energetic lepton.
This correlation has been shown to be useful
 in probing
CP violation in top production \cite{pes}.
Let us consider now the  $\hc b$ top decay mode.
If this decay is dominated by the  term proportional to $(1-\gamma_5)$
in (\ref{coup}), the $b$ will also be left-handed and
we will have a similar energy dependence as the $Wb$ decay mode.
Nevertheless, for $\tan\beta>1$ the term proportional to $(1+\gamma_5)$
cannot be neglected. This term involves a $b_R$ and therefore  the
relative energies of $b$ and $\hc$ (or their decay leptons)
are reversed: $t_L$ will produce an energetic lepton and $t_R$ a
less-energetic one.

In figs.~2 and  3 we show the lepton energy spectra for a
$t_L$ and $t_R$ respectively decaying through an on-shell $\hc$ or $W$.
We have taken $m_t=140$ GeV, $E_t=250$ GeV, $m_\hc= 100$ GeV and
 $\tan\beta=10$.
 In the $t_L$ decay, one can appreciate a significant
  difference  between the $\hc$ and $W$ curve. Such a difference is
smaller for $t_R$.
For small values of $\tan\beta$, the second term of (\ref{coup}) is small
 and the $\hc$ curve is shifted to the
left (right) for $t_L$ ($t_R$).

%\begin{figure}
%\plotpicture{\hsize}{8cm}{tl.topdraw}
%\vskip-4cm
%\epsfysize=16cm
%\centerline{\epsffile{tl.ps}}
%\vskip-4cm
%\vskip5pt
%\baselineskip=12pt
%\centerline{{\tenbf Figure 2:} \tenrm
%Lepton energy distribution for $t_L$ decaying through an on-shell
%$H^+$ or $W$.}

%\vskip15pt
%\plotpicture{\hsize}{8cm}{tr.topdraw}
%\vskip-4cm
%\epsfysize=16cm
%\centerline{\epsffile{tr.ps}}
%\vskip-4cm
%\vskip5pt
%\baselineskip=12pt
%\centerline{{\tenbf Figure 3:} \tenrm
%Lepton energy distribution for $t_R$ decaying through an on-shell
%$H^+$ or $W$.}
% \end{figure}

In future
hadron or $e^+e^-$ colliders the numbers of $t_L$ will be similar to that
of $t_R$ --unless we polarize the initial beams.  Therefore, the
 differences in the energy spectra of the decay
 leptons from $\hc$ and $W$ cannot be appreciated.
  Nevertheless, if the center of mass energy of the $\tt$ pair
is much larger that $2m_t$, we will have
\begin{equation}
N(t_L\bar t_R)\approx N(t_R\bar t_L)\gg N(t_L\bar t_L),\,
 N(t_R\bar t_R)\, .\label{ntt}
\end{equation}
This is the case of the $e^+e^-$ Next Linear Collider (NLC) where
$\sqrt s=500$ GeV or 1 TeV. Considering (\ref{ntt}),
 the strategy seems clear
for enhancing the sample of Higgs-mediated decays.  First, select $\tt$
events where one top decays to a $\tau$ and the other decays to a lepton
($\ell=e,\mu$).  For $\tt\ra W^+W^-$ events the energies of the lepton and
the $\tau$ will be comparable; both energetic for $t_R\bar t_L$ or both
non-energetic for $t_L\bar t_R$.  However, for $\tt\ra H^\pm W^\mp$ the
lepton and the $\tau$ will typically have different energies.  Thus,
we can enhance the $H^\pm W^\mp$ sample relative to the $W^+W^-$ by
selecting for events with $E(\tau)\gg E(\ell)$ or $E(\ell)\gg E(\tau)$.

To get an idea  of how the kinematic cuts can enhance the $H^\pm W^\mp$
signal,
we show in table 1 the number of $\ell-\tau$ pairs expected
 from $\tt\ra W^+W^-, H^\pm W^\mp\ra \ell\tau$  for an $e^+e^-$ collider
with $\sqrt s=500$ GeV and luminosity of 50 ${\rm fb}^{-1}$.  As before
we take $m_t=140$ GeV and $m_\hc=100$ GeV.
Since for $\tan\beta>1$,
  $\hc$ decays  into
$\tau\nu$ with a branching ratio close to one\footnote{\ninerm
\baselineskip=9pt QCD
 corrections \cite{men} reduce considerably
the $\hc\ra c\bar s$ branching
ratio and therefore
enhance BR($\hc\ra\tau\nu$).},
a clear signal of a charged Higgs from  a top decay will be the
 breaking  of lepton universality \cite{barn},
  \ie, an excess of $N_{\ell\tau}$ over $N_{\ell\ell}$.
We have computed $N^{WW}_{\ell\tau}\equiv N^{WW}_{\ell\ell}$
and $N^{HW}_{\ell\tau}$ without cuts, and with
$x_{\ell,\tau}\equiv 2E_{\ell,\tau}/E_t$
restricted to simultaneously lie between
 $1.2<x_\ell<2$ and $0<x_\tau<0.5$ or with $x_\ell$ and $x_\tau$
interchanged.  These rather strict cuts definitely increase the
percentage of $H^\pm W^\mp$ events, although at a considerable loss of
statistics.
In addition, for values of $\tan\beta$ that give
equal BR$(t\ra\hc b)$ the cuts increase the small differences in
$N^{HW}_{\ell\tau}$ that
arise from the dependence of
BR($\hc\ra\tau\nu$) on $\tan\beta$.

In this paper we have shown that the differences in the lepton energy
spectra from top decay to $H^+$ and $W^+$ may be useful in enhancing
the charged Higgs signal and for studying its couplings.  Using
correlations in the $\tt$ pairs (\ref{ntt}),
 we exhibited this difference by
making simple cuts on the lepton energies at the NLC to enhance the
Higgs signal.  It remains to be seen whether this effect can be used
at hadron colliders, where the $\tt$ correlations are presumably
weaker.  Further analysis is required in order to optimize the use of
the lepton energy information, especially in conjunction with other
techniques for enhancing the $H^+$ signal, such as $\tau$
polarization \cite{bul} or $b$ tagging \cite{barn}.  The lepton
energy spectra should be particularly useful at an NLC with polarized
beams, which can produce a large sample of longitudinally polarized top
quarks.

\begin{table}
\centerline{Table 1}
\vskip2pt
\begin{displaymath}
\vbox{\tabskip=0pt \offinterlineskip
\def\tablerule{\noalign{\hrule}}
\halign{\vphantom{\Big[}\strut#&\vrule#&\hfil #\hfil &&\vrule#&
\quad \hfil #\quad \cr
\tablerule
&&$\ \tan\beta$\ && 1.4 && 20 && 2.8 && 10 && 3.5 && 8 &\cr\tablerule
&&$\ {\rm BR}(t\ra\hc b)$\
&&\multispan3\hfil $1.4\cdot 10^{-1}$ \hfil
&&\multispan3\hfil $4.5\cdot 10^{-2}$ \hfil
&&\multispan3\hfil $3.3\cdot 10^{-2}$ \hfil&\cr\tablerule
&&$\ N^{WW}_{\ell\tau}$ (no cuts)\
&&\multispan3\hfil 1157 \hfil
&&\multispan3\hfil 1439 \hfil
&&\multispan3\hfil 1475 \hfil&\cr\tablerule
&&$\ N^{HW}_{\ell\tau}$ (no cuts)\
&& 1459 && 1736 && 603 && 610 && 453 && 455 &\cr\tablerule
&&$\ N^{WW}_{\ell\tau}$ (cuts)\
&&\multispan3\hfil 40 \hfil
&&\multispan3\hfil 50 \hfil
&&\multispan3\hfil 51 \hfil&\cr\tablerule
&&$\ N^{HW}_{\ell\tau}$ (cuts)\ && 106 && 185 && 46 && 63 && 36 && 46 &
\cr\tablerule}}
\end{displaymath}
\end{table}
\pagebreak
{\elevenbf\noindent References \hfil}
\vglue 0.4cm

\vglue 0.5cm

\end{document}